\documentclass{vgtc}                          % final (conference style)
\ifpdf%                                % if we use pdflatex
  \pdfoutput=1\relax                   % create PDFs from pdfLaTeX
  \usepackage{graphicx}                % allow us to embed graphics files
  \DeclareGraphicsExtensions{.pdf,.png,.jpg,.jpeg} % for pdflatex we expect .pdf, .png, or .jpg files
\else%                                 % else we use pure latex
  \usepackage{graphicx}                % allow us to embed graphics files
  \DeclareGraphicsExtensions{.eps}     % for pure latex we expect eps files
\fi%

\graphicspath{{figures/}{pictures/}{images/}{./}} % where to search for the images

\usepackage{microtype}                 % use micro-typography (slightly more compact, better to read)
\PassOptionsToPackage{warn}{textcomp}  % to address font issues with \textrightarrow
\usepackage{textcomp}                  % use better special symbols
\usepackage{mathptmx}                  % use matching math font
\usepackage{times}                     % we use Times as the main font
\usepackage{cite}                      % needed to automatically sort the references
\usepackage{tabu}                      % only used for the table example
\usepackage{booktabs}                  % only used for the table example
\onlineid{1083}

\vgtccategory{Research}

\vgtcinsertpkg

\usepackage{xspace}
\newcommand{\etal}{\emph{et al.}\@\xspace}
\newcommand{\ie}{\emph{i.e.}\xspace}
\newcommand{\eg}{\emph{e.g.}\xspace}
\newcommand{\etc}{\emph{etcetera}\@\xspace}

\newcommand{\systemName}{FairFuse\xspace}
\title{\systemName: Interactive Visual Support for Fair Consensus Ranking}

\author{Hilson Shrestha\thanks{e-mail:hshrestha@wpi.edu} %
\and Kathleen Cachel\thanks{e-mail:kcachel@wpi.edu} %
\and Mallak Alkhathlan\thanks{e-mail:malkhathlan@wpi.edu}
\and Elke Rundensteiner\thanks{e-mail:rundenst@wpi.edu}
\and Lane Harrison\thanks{e-mail:ltharrison@wpi.edu}}
\affiliation{\scriptsize Worcester Polytechnic Institute}
\teaser{
  \centering
  \includegraphics[width=\linewidth]{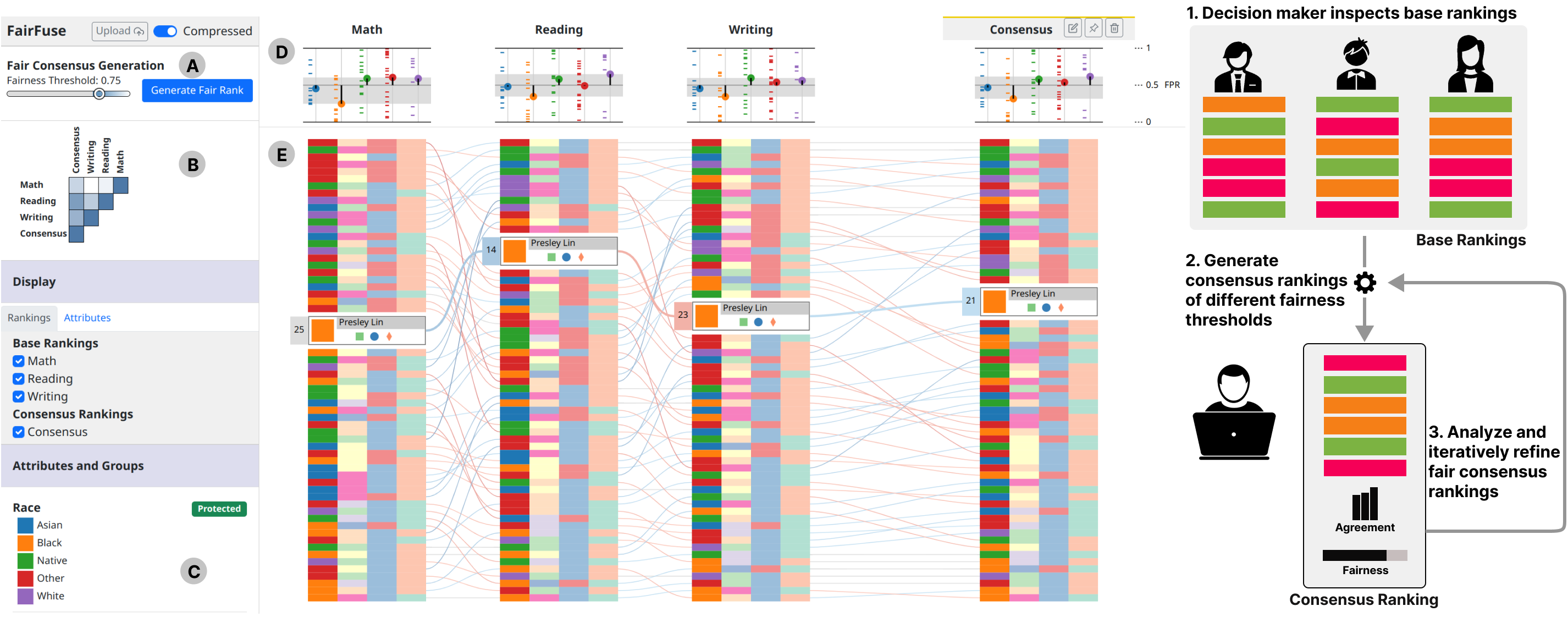}
  \caption{\systemName provides multiple views supporting fairness-oriented ranking workflows: A) Consensus Generation View, B) Rank Similarity View, C) Attribute / Protected Attribute Legends, D) Group Fairness View, E) Ranking Exploration View. (Right) Illustrating a fairness-oriented ranking workflow enabled by \systemName.}
  \label{fig:teaser}
}

\abstract{
Fair consensus building combines the preferences of multiple rankers into a single consensus ranking, while ensuring any group defined by a protected attribute (such as race or gender) is not disadvantaged compared to other groups. 
Manually generating a fair consensus ranking is time-consuming and impractical--- even for a fairly small number of candidates. 
While algorithmic approaches for auditing and generating fair consensus rankings have been developed, these have not been operationalized in interactive systems. 
To bridge this gap, we introduce \systemName, a visualization system for generating, analyzing, and auditing fair consensus rankings. 
We construct a data model which includes base rankings entered by rankers, augmented with measures of group fairness, and algorithms for generating consensus rankings with varying degrees of fairness.
We design novel visualizations that encode these measures in a parallel-coordinates style rank visualization, with interactions for generating and exploring fair consensus rankings. 
We describe use cases in which \systemName supports a decision-maker in ranking scenarios in which fairness is important, and discuss emerging challenges for future efforts supporting fairness-oriented rank analysis.
Code and demo videos available at \url{https://osf.io/hd639/}.
} % end of abstract

\CCScatlist{
  \CCScatTwelve{Human-centered computing}{Visualization}{Visualization systems and tools}{}
}

\begin{document}

%% The ``\maketitle'' command must be the first command after the
%% ``\begin{document}'' command. It prepares and prints the title block.

%% the only exception to this rule is the \firstsection command
\firstsection{Introduction}

\maketitle

%% \section{Introduction} %for journal use above \firstsection{..} instead
The ubiquitous task of combining preferences by multiple stakeholders into a consensus is challenging for decision-makers that steer this process. Decision-makers often grapple with diverging preferences provided by different stakeholders, and must reach a single decision that all stakeholders accept and agree with. A frequent approach to such decision-making is to employ rankings, where each stakeholder provides their ranking over the candidates. Candidates might include lists of people, organizations, or other entities. Decision-makers combine these base rankings from individual stakeholders into a single consensus ranking as part of the process. 

However, when ranking candidates, stakeholders may provide biased or unfair rankings \cite{chouldechova2020snapshot}. Bias can be implicit (unintended), for example, when favoring candidates from a particular university who happen to be overwhelmingly white. 
Bias can also be explicit, for example, weighing women candidates lower due to a perceived lack of ability for the target role. 
%Bias in a given base ranking can even be relatively small and limited to a few candidates, but when combined into an aggregate consensus ranking, the result risks amplifying or perpetuating systemic bias of certain socio-demographic groups over others. 
One way to mitigate such unfair outcomes is by promoting measures from the algorithmic fairness community, such as \emph{group fairness} or \emph{statistical parity} \cite{pedreshi2008discrimination}. 
Statistical parity, for example, is a requirement that all groups receive an equal proportion of the positive outcome; in our case, favorable positions in the consensus ranking. 
Without intervention in the ranking process, there is substantial risk of perpetuating unfair practices, and thus harming marginalized groups.

Unfortunately, constructing a  consensus ranking is challenging \cite{bartholdi1989voting, dwork2001rank} and ensuring that this consensus ranking  is fair is even more difficult \cite{kuhlman2020rank, cachel2022manirank}. 
Numerous visualization tools have explored the design space of rankings  \cite{gratzl2013lineup} and rank-based decision making \cite{mahyar2017consesnsus, hindalong2020towards}. 
But existing approaches have not dealt with the complications of incorporating fairness into visual encodings, nor with interactive workflows related to consensus rank generation. Similarly, while research in fair algorithms has developed rank-focused auditing metrics and fair rank aggregation methods \cite{kuhlman2020rank, cachel2022manirank}, they have been confined to (non-visual) algorithmic solutions requiring substantial technical expertise to use.

% Our approach
To address this gap, we contribute the design and development of {\bf \systemName}, an  interactive visualization system for 
generating, analyzing, and auditing fair consensus rankings. We develop a model capturing rankings and candidate attributes for identifying candidate groups, group-based fairness metrics, and algorithms to generate fair consensus rankings.
We propose a parallel-coordinates style visualization design for rankings with a focus on the group membership of candidate attributes. We develop novel visual encodings for group-based fairness metrics. 
\systemName enables an iterative ranking- and fairness-oriented workflow,
allowing decision-makers to  visually inspect and edit consensus rankings as part of their decision-making process. Our use cases demonstrate how a decision-maker can use \systemName in fairness-oriented ranking scenarios. We conclude by discussing  emerging challenges in supporting fairness in ranking-based decision-making through interactive visualization systems. 

\section{Related Work}

Visualization systems have been designed to aid decision-makers in inspecting 
stakeholder preferences for  decision-making tasks \cite{bajracharya2018interactive, carenini2004valuecharts, dimara2017dcpairs, hansen2008new, hayez2012d, hong2018collaborative,liu2018consensus, pham2016qstack, weng2018srvis, weng2018homefinder, shah2014collaborative, mustajoki2000web}.
Some consider settings, like ours, in which  preferences  from multiple stakeholders
are modelled as rankings \cite{hindalong2020towards, liu2018consensus, hayez2012d, hansen2008new}.
Most recently, Hindalong \etal \cite{hindalong2022abstractions} developed visual abstractions for inspecting and comparing two or more  preferences.
However,  these works neither apply algorithms
to automatically  construct  one integrated fair ranking 
nor does their integration of multiple rankings address  the critical real-world challenge   that  preferences tend to contain biases about socio-demographic groups (\eg different gender or racial identities). 
%Unfortunately,these works thus neither  provide the much needed bias inspection nor bias mitigation solutions.

While interactive and visual systems \cite{bantilan2018themis, google_whatif, ai_360, saleiro2018aequitas, bird2020fairlearn, xie2021fairrankvis, yang2018nutritional, ahn2019fairsight, jin2017collaborative}  highlight and mitigate socio-demographic biases, they are not targeted towards rank-oriented workflows and fair consensus rankings. 
Instead, they target predictive machine learning tasks like classification \cite{bantilan2018themis, google_whatif, ai_360, saleiro2018aequitas, bird2020fairlearn, xie2021fairrankvis} or restrict themselves to single ranks \cite{yang2018nutritional, ahn2019fairsight}. 

In the algorithmic fairness community,
the predominant mitigation to bias and discrimination is the notion of ``group fairness'' \cite{pedreshi2008discrimination}.
%This aims to ensure decision outcomes are not unfavorable to specific socio-demographic groups. 
Group fairness is conceptualized as treating groups similarly \cite{pedreshi2008discrimination}. The state-of-the-art includes both {\it metrics} \cite{kuhlman2019fare, geyik2019fairness, beutel2019fairness, narasimhan2020pairwise, singh2018fairness, yang2017measuring} to quantify bias in rankings and {\it algorithms} \cite{kuhlman2020rank} to generate such fair consensus rankings. Specifically, Kuhlman \etal \cite{kuhlman2020rank} address fair-consensus ranking generation for two socio-demographic groups, while Cachel \etal \cite{cachel2022manirank} extend this scope to the multi-group setting. 
\systemName takes initial steps towards leveraging
these algorithmic solutions and their metrics to aid decision-makers in combining multiple stakeholder rankings into fair consensus rankings.

\section{Basics of Fair Consensus Ranking}

We characterize the data model and tasks for decision-makers analyzing multiple stakeholder preferences and ultimately combining them to generate a fair consensus ranking. 
%Additionally, we describe the data we use in illustrating how \systemName works.

\subsection{Abstraction of Data Model}

%In providing decision-makers with support for exploring and then combining preferences from multiple stakeholders into a single consensus decision, we construct the  data model below.

We are given a set of \textbf{candidates}, described by \textbf{attributes}, to be ranked. One of the attributes, typically a categorical attribute referred to as the \textbf{protected attribute} (such as gender, race, or income-level), is associated with bias measurement and mitigation. We refer to candidates sharing the same value of the protected attribute as {\bf groups}, such as Man, Woman, or Non-binary groups in the Gender attribute.
Stakeholders in the committee  (called {\bf rankers})   each order (rank) the set of candidates to create a list of {\bf base rankings} provided to the {\bf decision-maker}. A decision-maker (head of the committee) using our system generates {\bf consensus rankings} with the aim to order the candidates such that the base rankings, and thereby rankers, mostly agree with the consensus ranking.

The consensus ranking also must be fair. For auditing the fairness of rankings, we employ two metrics: a
group-specific pairwise fairness metric \textbf{FPR} (Favored Pair Representation) \cite{cachel2022manirank} to measure the fair treatment of each group in the  ranking, and an aggregate fairness metric \textbf{ARP} (Attribute Rank Parity) \cite{cachel2022manirank} to quantify if the overall ranking
across all groups satisfies the statistical parity fairness criteria \cite{pedreshi2008discrimination}. In generating a fair consensus ranking, the decision-maker sets the {\bf fairness threshold} value which controls the level of ARP represented in the consensus ranking. The later is then generated by a function utilizing the \textbf{Fair-Copeland Algorithm} \cite{cachel2022manirank}.  A function \textbf{Kendall Tau distance} \cite{kendall1938new} computes the similarity/agreement between any two rankings.

\subsection{Task Analysis}

We define nine abstract tasks to guide the development of FairFuse, following procedures from task abstraction methodologies such as Lam \etal \cite{lam2017bridging}, and recent work on group decision making from Hindalong \etal \cite{hindalong2020towards}. 
These tasks support comparing rankings, investigating bias in rankings, and iteratively generating fair consensus rankings.

\textbf{T1: Identify candidate positions} across base rankings to assess high and low performing candidates according to the rankers.

\textbf{T2: Identify protected attribute values} (\ie group membership of) and additional attributes of ranked candidates.

\textbf{T3: Analyze the (dis)similarity across rankings}, both between base rankings  and between base and consensus rankings.  

\textbf{T4: Explore the distribution of the placement of groups}, in each ranking, to compare advantages across groups.

\textbf{T5: Understand the fair treatment or lack thereof of each group} per ranking, as captured by the FPR metric.

\textbf{T6: Intuit fairness of each ranking} as a whole with respect to the statistical parity fairness criteria for the specified protected attribute, as captured by the ARP metric.

\textbf{T7: Compare group fair treatment and fairness} across rankings, both base and consensus alike.

\textbf{T8: Generate consensus rankings} by initiating the generation algorithm, while controlling their level of fairness.

\textbf{T9: Iterate on and adjust generated consensus rankings} to satisfy the desired trade-off between base rankings and degree of fairness.

\subsection{Data Sets for Use Case Scenarios}

For the demonstration of this fair ranking problem, we use a
dataset of 60 students with scores in 3 subjects: Math, Reading and Writing\cite{kimmons}, and convert them into base rankings.
% To generate three base rankings, we use exam scores in the above three subjects. 
We use the provided race attribute composed of 5 abstract groups (Group A, Group B, ...) as a protected attribute and map it to concrete race categories: White, Black, Asian, \etc.
% For protected attributes, we use the provided race attribute which
% is composed of  5 abstract groups (Group A, Group B, ...).
% To illustrate the use of \systemName, we map groups to concrete race categories: White, Black, Asian, \etc. 
Since names are not provided,  we generate random names for each candidate. 
While this dataset is used to demonstrate the visual encodings and features, \systemName supports other datasets--- an example use case with employee bonus distributions data is included in the supplement.

\section{\systemName Overview}

\systemName is designed to support the process of both analyzing and combining preferences from multiple rankers into a {\it fair} consensus ranking. 
In designing \systemName, we develop core views based on parallel coordinates, augmented with custom visual encodings for fairness metrics, and interactive components for generating fair consensus rankings.

\subsection{Ranking Exploration View}

\begin{figure}[tb]
  \centering
  \includegraphics[width=\columnwidth]{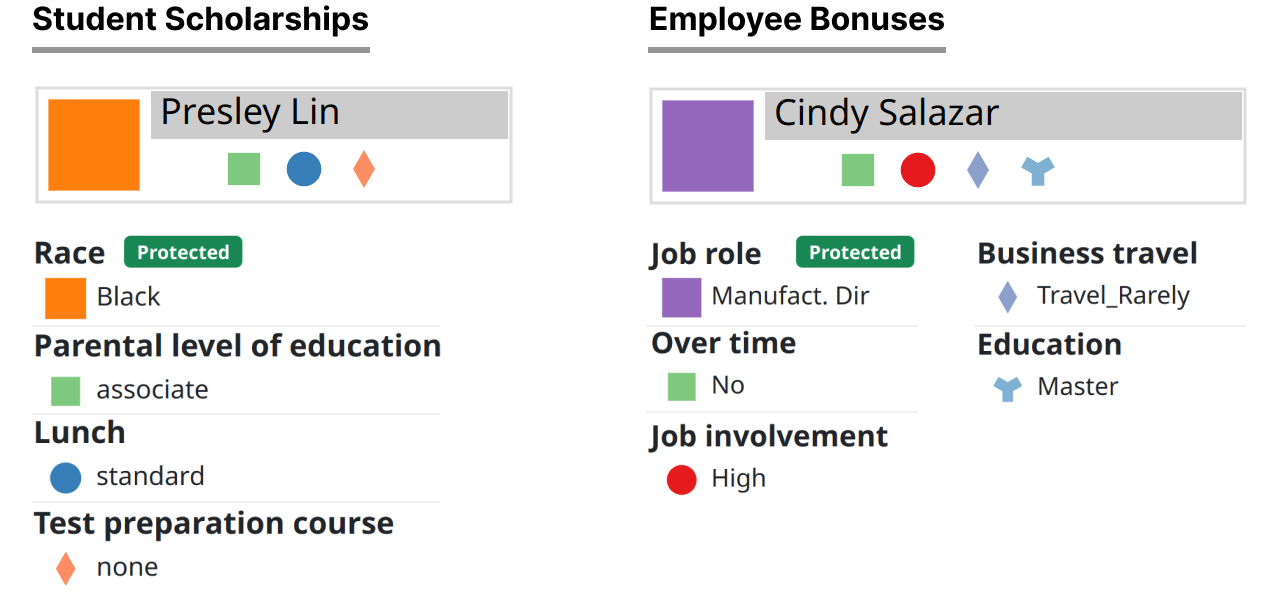}
  \caption{Candidate Card contains glyphs representing multi-variate attributes. %Shown are  cards for two scenarios.
  The protected attribute is emphasized with a large (versus smaller) shape.  Color represents the value of the attribute.}
  \label{fig:attribute-glyphs}
\end{figure}

The {\em{Ranking Exploration View}} (Figure \ref{fig:teaser}E) contains all  candidates
ordered into two or more base rankings ({\bf T1}). Columns on the left correspond to input base rankings; while fair consensus rankings generated by the decision maker are appended to the right upon their creation. 
Drawing on designs from Nobre \etal \cite{nobre2020evaluating} and Maguire \etal \cite{maguire2012taxonomy}, candidate attributes  are displayed as glyphs in a {\em{Candidate Card}} (Figure \ref{fig:attribute-glyphs}) ({\bf T2}).

This view uses parallel coordinates to compare candidates across different rankings ({\bf T3}), drawing on features from similar rank oriented systems such as LineUp \cite{gratzl2013lineup} and Hindalong \etal \cite{hindalong2020towards}. 
The order of candidates in a given column is based on candidate rank in the case of a base ranking columns, or the Fair-Copeland Algorithm in generated rankings.
Each candidate appears across all rankings, with lines connecting them to illustrate change in position across rankings.
Lines connecting the candidate across the rankings are colored based on the degree of change in the candidate's position between adjacent rankers. 
Candidates ranked higher in the subsequent ranking are colored in a gradient scale of blue, while those  ranked lower  are colored in a gradient scale of red.
%However,  when  a large number of rankers and a significant disagreement amongst them exists, a parallel-coordinates plot can result in significant clutter \cite{heinrich2013state}. 

\begin{figure}[t]
  \centering
  \includegraphics[width=\columnwidth]{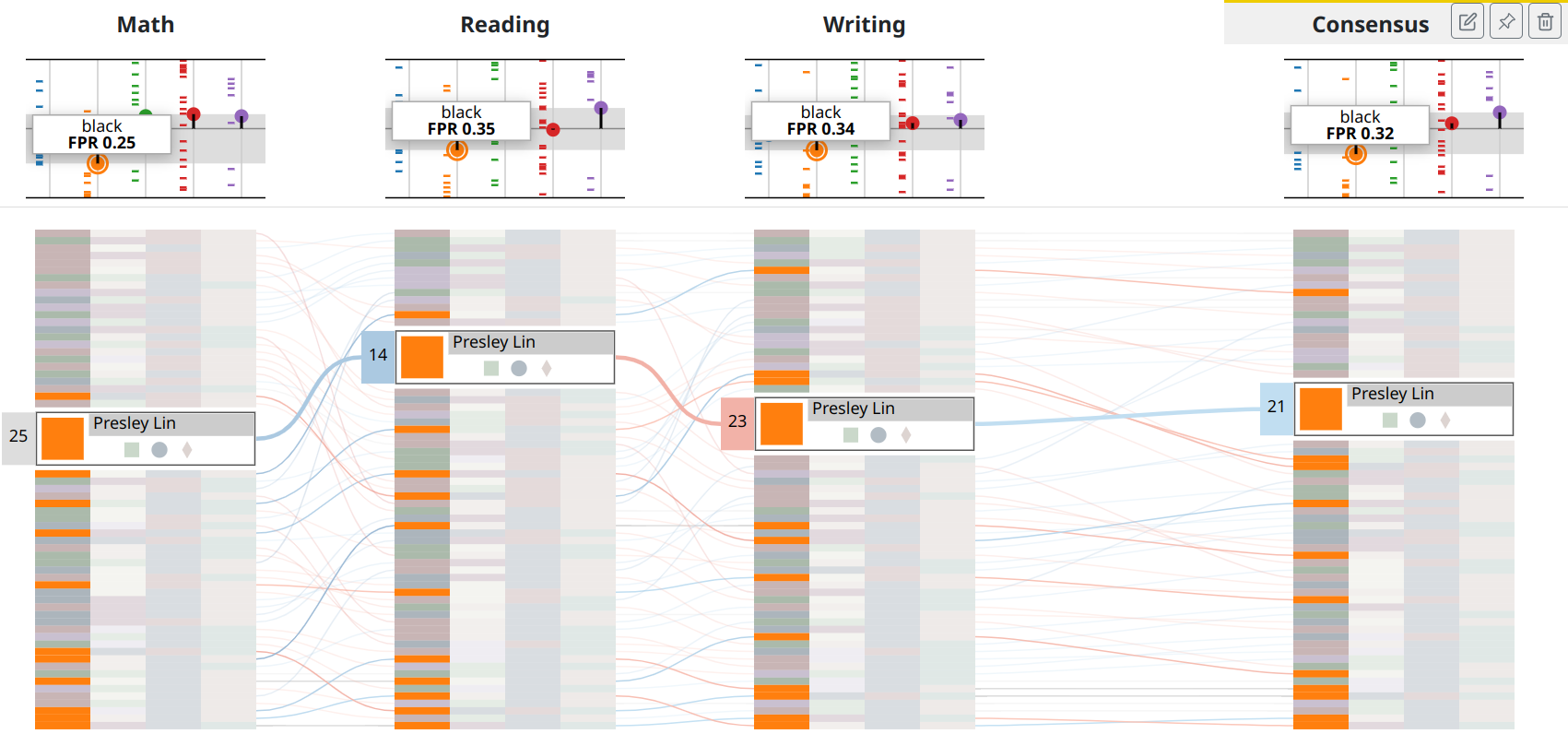}
  \caption{Compressed view with a group selected. When a group is selected in the Group Fairness View, all candidates of that group are highlighted, facilitating the decision-maker in focusing on both group and individual fairness considerations.}
  \label{fig:inspectupdate}
\end{figure}

To reduce parallel coordinates clutter (\eg \cite{heinrich2013state}) while maintaining task effectiveness, we hide lines for which both candidates on adjacent rankers are not visible within the screen.
Clutter can also result from orderings of parallel coordinate columns \cite{blumenschein2020evaluating}. 
Users can drag to re-arrange columns, and \systemName can be readily extended with automatic ordering techniques.
We also design a {\em Compressed {Ranking View}} mode (Figure \ref{fig:inspectupdate}) which represents a scaled-down version of the rankings. In this mode, the candidate cards (Figure \ref{fig:attribute-glyphs}) are initially hidden, but appear when hovering over a particular candidate. 
The protected attribute glyph is displayed with full saturation so that the decision-maker can explore how groups are distributed in each ranking ({\bf T4}), while other attributes are desaturated so as to be visible while interfering less with the protected attribute color.

\subsection{Group Fairness View}

To support auditing rankings in terms of fairness,
the {\em{Group Fairness View}} (Figure \ref{fig:teaser}D) compactly captures fairness of a ranking at multiple levels of granularity: both at the level of individual groups, and holistically across groups for assessing fairness across rankings. 
The FPR metric \cite{cachel2022manirank} captures if a specific group 
is fairly treated throughout the ranking ({\bf T5}). 
Specifically, FRP score $= 0.5$ denotes totally fair group treatment, while $< 0.5$ represents under-advantage and $> 0.5$ over-advantage. 
The ARP metric \cite{cachel2022manirank} captures if statistical parity fairness is satisfied  by the ranking overall ({\bf T6}), \ie, {\it all} groups are comparably treated to each other. Here, ARP $= 0$ is absolute fairness, anything higher is further and further from total fairness. 
This novel fairness view is critical to capture the notion that  in multiple-group settings one or more groups may be fairly treated, while others may be unfairly over- or under-advantaged.
%This view allows the decision-maker to visually assess variances in group treatments and understand how close a ranking is to balancing fairness across groups. 

\begin{figure}[tb]
  \centering
  \includegraphics[width=\columnwidth]{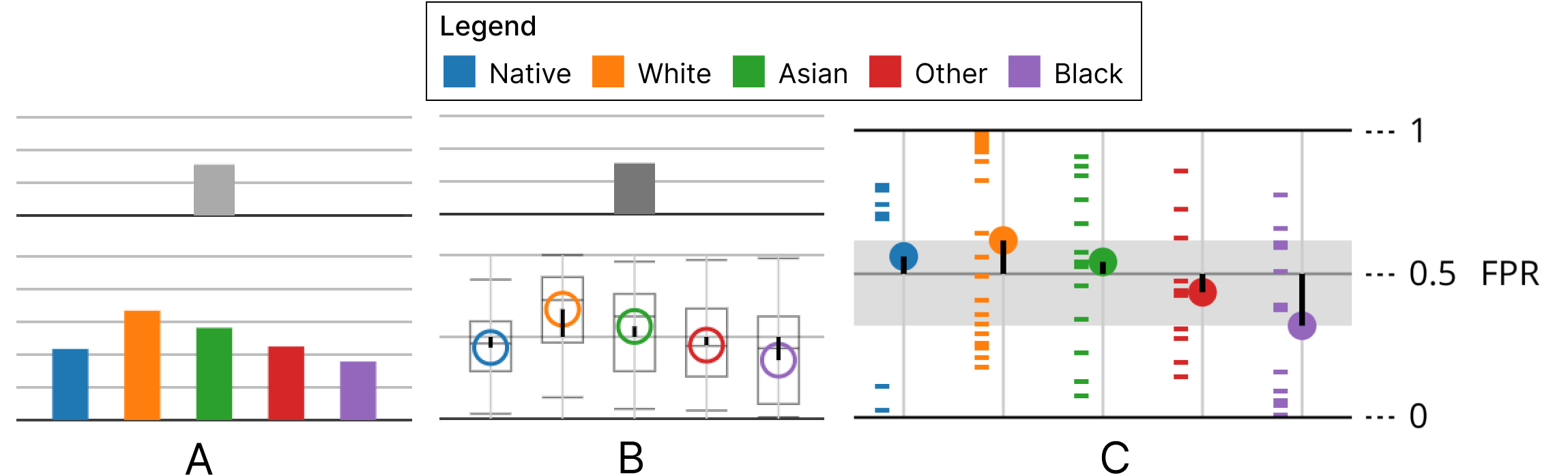}
  \caption{Group Fairness View design iterations. A) Colored barchart for FPR; gray bar for ARP. B) Dot plot for FPR, box-plot for distribution of groups in ranking. C) Final Design: Dot plot represents FPR, heatmap the distribution of groups, and shaded region the ARP.}
  \label{fig:fairness-metric-view}
\end{figure}

In designing the Group Fairness View, we initially explored 2 alternate prototypes (Figure \ref{fig:fairness-metric-view}). 
Because FPR and APR are scalar values, we first represented the FPR and ARP fairness scores with bar encodings at the top of ranking columns (Figure \ref{fig:fairness-metric-view}A). 
However, after determining that 
this design made it  difficult to identify over-advantaged and under-advantaged groups ({\bf T5}), 
our second design  placed an axis at FPR $0.5$ and adopted a hybrid dot-plot and box-plot encoding (Figure \ref{fig:fairness-metric-view}B).
This change supports more semantically meaningful visual queries.
For example, dots above FPR $0.5$ represent over-advantaged groups, informing that they are unfairly receiving a larger share of favorable rank positions 
({\bf T5}).

As explained below, 
our  third and final design variation of the Group Fairness View as depicted in Figure \ref{fig:fairness-metric-view}C offers additional advantages. 
Since ARP measures the difference between the maximum and minimum FPR scores, we can visualize the ARP score with the region between 
the scores of the respective group. 
This change enables visual queries within and across rankings to assess group fairness of each ranking({\bf T6}), as a smaller ARP value would create a smaller shaded region.
To mitigate the limitations of boxplots for showing non-contiguous distributions, we adopt a marginal mark-based distribution plot.
Finally, the view affords interactive features, such as displaying exact FPR and ARP values on hover, and highlighting groups in the parallel coordinates plot on click.

\subsection{Consensus Generation and Similarity View}

The {\em{Consensus Generation and  Similarity View}} (Figure \ref{fig:teaser}A) provides functionality for the decision maker to generate and compare fair consensus rankings ({\bf{T8}}). 
The consensus generation component includes a fairness threshold slider used to adjust the level of fairness that should be reflected in the generated consensus ranking. 
Algorithmically, this is accomplished by passing the base rankings and
the fairness target value to  the recently innovated Fair-Copeland algorithm \cite{cachel2022manirank}, which then computes and returns a new consensus ranking.
Because a set of base rankings are unlikely to be completely unfair from the outset, the slider includes a gradient overlay to indicate that the fairness threshold will only produce fairer results if changed in a particular region. 
Similarly, on the other extreme, if the slider is set to 0, it will generate a consensus ranking  soley based on the  input base rankings.

As the decision maker initiates the generation of a consensus ranking,
they can assess the overall agreement between the base versus this new consensus rankings (\textbf{T3}) via the  similarity matrix view (Figure \ref{fig:teaser}B).
For example, if they  generate rankings with a high level of fairness, the resulting ranking may deviate more from some base rankings than others.
Similarity is calculated 
using a common measure for rank dissimilarity called Kendall-Tau distance \cite{kendall1938new}, with
darker squares representing more similarity between two rankings.
This similarity component also aids the decision-maker in iterating over alternate consensus rankings ({\bf T9})  to finalize the consensus decision. 
In designing this view, we considered alternatives such as arc diagrams embedded
into  the 
ranking exploration view (which were too cluttered), dissimilarity as the metric was traditionally defined (with inversion considered more interpretable), and variations of the matrix orientation.

\subsection{Additional Interactions and Workflow}

\systemName provides additional interactions to support the decision-maker in a fairness-oriented rank analysis and generation workflow. Ranking Exploration, Group Fairness, and Consensus Generation and Similarity Views  include design elements that respond to user actions such as clicks and hovers. A user hovering in the Ranking Exploration view, for example, will highlight a Candidate Card for easier exploration across views (\textbf{T1, T2}). A click in this scenario  ``pins'' a candidate for comparison against other candidates. Brushing is also enabled \cite{heinrich2013state}, allowing the user to drag  select ranges of candidates within particular columns, which is particularly useful in the compressed views. Similar hover and click functions are available in other views, mainly oriented towards emphasizing or de-emphasizing visual components to enable the decision-maker to focus on particular  tasks.

To support iteration and adjustment of consensus rankings (\textbf{T9}), \systemName provides the decision-maker with editing features on consensus rankings.
 \systemName  supports manual editing of fair consensus rankings (\textbf{T7}), as the decision-maker may have additional context and information that they need to preserve in the resulting ranking.
Decision-makers may adjust the fairness threshold of a consensus ranking to obtain another result, create or ``pin'' rankings, and manually adjust the position of candidates.
Importantly, repositioning candidates immediately triggers the recalculation of fairness metrics, showing the decision-maker how fairness is lost or gained through their manual editing.

\section{Use Case Scenarios Using the \systemName System}

A scholarship administrator, Jo, is responsible for determining the merit scholarship package of prospective students\footnote{An additional usage scenario is presented in supplemental material.}.
Jo needs to combine the recommendations of three rankers, teachers in Math, Reading, and Writing, and form a single ranking to allocate the merit scholarships.
Cognizant that systemic and societal biases can affect how students of differing races perform in academic subject exams, which in turn can affect how students are perceived by subject-specific rankers, Jo seeks to detect and mitigate excess bias in the consensus ranking to ensure all groups are comparably treated. 

Jo loads the data of base rankings given by the teachers along with candidate attribute information into the \systemName. 
Jo uses the \textit{Similarity View} to assess to what degree each base ranking agrees with others, along with visually inspecting the lines between adjacent rankings in the \textit{Rank Exploration View}. 
At this point Jo uncovers that the Math teacher disagrees to some extent with the other teachers, which can make a consensus challenging, even before considering fairness. 

Next, Jo switches to the compressed view to evaluate how candidate attributes are distributed across rankings. 
Auditing primarily for fairness, however, Jo pays particular attention to the protected attribute, race.
For this task, Jo studies the \textit{Group Fairness View} on top of each ranking (Figure \ref{fig:teaser}D), which shows distributions of protected attributes throughout the ranking.
Jo notices immediately that the FPR fairness metric indicates that white students have a stronger advantage over students from other races.
On closer examination, Jo discovers that across all rankings, students from the white group are clustered at the top, while students from the black group are clustered more towards the bottom. 
This is then reflected in the ARP scores (gray area) of the rankings, indicating the base rankings in general are far from fair as defined by statistical parity.

After exploring and comparing the similarities and fairness of the base rankings (Figure \ref{fig:teaser}B,D), Jo initiates the auto-generation of a consensus ranking, using the Consensus Generation View. 
Immediately, Jo notices that the consensus ranking reflects the biases found in base rankings. 
Jo then progressively adjusts the Fairness Threshold (Figure \ref{fig:teaser}A) to generate a fairer consensus ranking. 
Throughout this process, Jo references the Similarity View matrix and base rankings themselves to evaluate the extent to which base rankings are represented in the fair consensus.
Honing in on a consensus ranking that balances the desired trade-off between the fairness and preference representation, Jo makes manual swaps between candidates to refine the target consensus ranking. 
With each edit, Jo's changes are audited visually by changes in the Group Fairness View (Figure \ref{fig:teaser}D), helping ensure this manual manipulation does not drastically change the desired fairness measure. 
The resulting consensus ranking is both fair with respect to mitigating the over-advantage of white students and their disproportionately large merit awards, while ensuring the teacher recommendations expressed by base rankings are adequately combined and represented.

\section{Discussion \& Future Work}
In designing \systemName, 
we learn about the challenges and opportunities for integrating fairness-oriented algorithms into visualization-driven workflows.
State-of-the-art algorithms only consider a single protected attribute per candidate, but future work should engage with 
the algorithmic and visualization challenges surrounding multiple protected attributes and intersectionality \cite{crenshaw1990mapping}.
The current similarity view shows the agreement between any two rankings, but matrix views can be confusing \cite{nobre2020evaluating}. 
Future designs might explore how particular aspects of the workflow can be used to inform new encodings that better represent agreement between base rankings and generated consensus rankings.
For the current implementation, we use the Fair-Copeland algorithm. Future work could also explore how to support multiple algorithms in a single tool, and comparisons between them.
While our use cases illustrate the utility of a system \cite{lam2011empirical}, task-based user studies involving fairness will also be important future work.
User studies might explore ethical issues surrounding algorithmic fairness, such as the potential to ``fair wash'' results by deceptively using metrics to promote unfair outcomes \cite{bietti2020ethics}.

\section{Conclusion}
Fusing preferences of multiple stakeholders into a fair consensus decision expressed by a result ranking is ubiquitous yet an incredibly challenging process. 
To support fair consensus-building workflows, we introduce \systemName, a visualization system for auditing, analyzing and generating consensus rankings. 
\systemName interactively aids the decision-maker in generating and refining fair consensus rankings given a set of base rankings. 
With custom visualizations encoding fairness metrics from fair-algorithms research, \systemName enables decision-makers to visually and interactively explore and audit  base- and generated- rankings.
We demonstrate how \systemName supports fairness-oriented ranking workflows through use cases, yielding a foundation for future studies at the intersection of fairness, ranking, and visualization.

\acknowledgments{
This work was supported in part by NSF IIS \#2007932.}

\bibliographystyle{abbrv-doi}

\bibliography{references}
\end{document}

\maketitle

\section{Sup-1: Datasets}
\begin{table*}[th!]

\begin{tabular}{p{0.15\linewidth} | p{0.1\linewidth} | p{0.65\linewidth}}
\hline
{\bf Name} & {\bf Symbol} & {\bf Representation} \\ \hline
Candidates & $X = x_1,...,x_n$ & $n$ candidates ranked \\ \hline
Base rankings & $R = r_1,…, r_m$ & $m$ rankings produced by multiple   stakeholders (rankers) \\ \hline
Protected Attribute & $p$ & Attribute for bias measurement/mitigation such as gender \\ \hline
Favored Pair Representation \cite{cachel2022manirank} & $FPR(G_{p:v}, r)$ & Measure quantifying if a group $G_{p:v}$ is fairly treated in ranking $r$. Ranges from $[0,1]$ where value $0.5$ denotes fair treatment, $0$ denotes complete disadvantage, and $1$ denotes over-advantage. \\ \hline
Attribute Rank Parity \cite{cachel2022manirank}& $ARP(p, r)$ & Measure quantifying if ranking $r$ is fair with respect to statistical parity for $p$. Ranges from $[0,1]$, $0$ denotes absolute fairness, and $1$ denotes   complete unfairness. \\ \hline
Kendall-Tau \cite{kendall1938new} & $distKT(r1,r2)$ & Ranking similarity metric comparing rankings $r_1$, and $r_2$. The value represents the number of pairs swap needed to convert $r_1$ to $r_2$. \\ \hline
Consensus Ranking & $C$ & Ranking that is closest to the base rankings, thereby best representing $R$. \\ \hline
Fairness Threshold & $t$ & Parameter controlling the maximum ARP of the fair consensus ranking, i.e., maximum resulting $ARP$ of fair consensus ranking = $1-t$ 
\\ \hline
Fair-Copeland Algorithm\cite{cachel2022manirank}  & $FC(R, X, p, t)$ & Method to generate a fair consensus ranking, such that all preferences in $R$ are maximally represented subject to the result having at most $ARP$ of $1 - t$ \\ \hline
\end{tabular}
\caption{Data model used for fair consensus ranking generation}
\label{tab.data}
\end{table*}

For the demonstration of generating fair consensus rankings, we used the following publicly available datasets for two different scenarios:
\begin{enumerate}
    \item Scholarships Distribution scenario: We used the dataset of 60 students with students scores in 3 subjects: Maths, Reading and Writing. The dataset contains multiple attributes but we utilize the exam scores in the dataset to convert it into 3 base rankings and build a fair consensus rankings by taking race as a protected attribute. Within the race attribute, there are 5 groups. \url{http://roycekimmons.com/tools/generated_data/exams}
    \item Employee Bonus Distribution scenario: We used a subset of dataset of Employee Attrition and Performance. We utilized the performance rating in the dataset to convert it into 3 base rankings (R1, R2 and R3). We randomized the rankings between these rankings so that they resemble the disagreement in rankings as in real-world. \url{https://www.kaggle.com/datasets/pavansubhasht/ibm-hr-analytics-attrition-dataset}
\end{enumerate}
These datasets do not contain individual names. So, we used the names python package \footnote{https://pypi.org/project/names/} for generating names.

\section{Sup-2: Detail of Data Model}

\subsection{Input}

Our system \systemName is designed to aid consensus decision-making in regards to a set of candidates for the task at hand (e.g., candidates for scholarships). We assume that in the set of $n$ candidates $X$, each candidate is described by a set of attributes $\mathcal{A}$, and categorical {\bf protected attribute} $p$ such as gender, race, or income-level.
We refer to $p$ as the protected attribute chosen by the user for bias measurement and mitigation. 
For each possible value $v$ of the protected attribute $p$, there is a {\bf group $G_{p:v}$} composed of individuals in set $X$ that have the same value $v$ for the protected attribute $p$. 
For instance, $G_{race:asian}$ is the group of all candidates that have the value asian for the race protected attribute.

The decision-maker utilizes our system to explore and combine (into a consensus decision) the preferences over $X$ of multiple stakeholders. We refer to these stakeholders as as {\bf rankers}. Each of the $m$ rankers provides a ranking $r_{i}$ ordering the candidates in $X$. We collectively refer to the rankings produced by the rankers as a set of {\bf base rankings}, $R = r_{1},..., r_{m}$.

\subsection{Metrics}

In quantifying bias in the base rankings $R$ our system employs the group fairness notion of {\bf statistical parity} \cite{pedreshi2008discrimination}. Statistical parity is a contemporary fairness notion stipulating that candidates must receive an equal proportion of the positive outcome regardless of their protected attribute value. In our setting, the positive outcome is favorable rank positions. Thus, when assessing statistical parity we measure if in the given ranking (a base ranking or consensus ranking) groups have equal proportions of favorable rank positions. To quantify statistical parity fairness we use the metrics introduced in Cachel et al. \cite{cachel2022manirank}. Namely, we utilize the metrics of $FPR$ (Favored Pair Representation) and $ARP$ (Attribute Rank Parity). For auditing rankings, the former quantifies how fairly a specified group is treated, while the later quantifies the presence (or lack) of statistical parity.

$ARP$ and $FPR$ are pairwise metrics. Any ranking  $r$ over $n$ candidates can be decomposed into (n(n - 1))/{2} pairs of candidates ($x_i, x_j$) where $x_i \prec_{r} x_j$. Specifically, the metrics count mixed pairs, where a mixed pair is pair comparing candidates from two different groups. For instance, a pair with two woman candidates is not a mixed pair, while a pair with a man and woman is a mixed pair. Intuitively, the more mixed pairs a group is favored in (or "wins"), the higher up in the ranking that group is compared to other groups.

The $FPR$ metric ranges from $[0,1]$, where a value of $0.5$ indicates the given group in the specified ranking is fairly treated \cite{cachel2022manirank}. A value of $0$ indicates the group is totally disadvantaged (i.e, occupying the bottom and thus worst rank positions), and a value of $1$ indicates the group is over advantaged (i.e, occupying the top and thus best rank positions). Equation \ref{eq.fpr} shows the calculation for the $FPR$ measure for a given group $G_{p:v}$ in ranking $r$, where $\omega_{M}(G_{p:v}, r)$ is the total number of mixed pairs in ranking $r$.

{
\begin{equation}
    FPR(G_{p:v}, r) = \sum_{x_i \in G_{p:v}} \sum_{x_l \notin G_{p:v}}\frac{countpairs(x_i \prec x_l)}{\omega_{M}(G_{p:v}, r)}
    \label{eq.fpr}
\end{equation}
}

Next, the $ARP$ measure utilizes the $FPR$ scores for all the groups in order to quantify if the given ranking satisfies statistical parity. Specifically the $ARP$ measure is the maximum absolute difference between $FPR$ scores. The $ARP$ measure ranges from $[0,1]$, when $ARP = 1$ then the protected attribute is maximally far from statistical parity and the ranking is maximally unfair \cite{cachel2022manirank}. Meaning, one group is entirely at the top of the ranking, while a second group is entirely at the bottom of the ranking.
When $ARP = 0$, perfect statistical parity is achieved and the ranking is totally fair. Equation \ref{eq.arp} shows the calculation for the $ARP$ measure for a given protected attribute $p$ in ranking $r$.

{
\begin{equation}
    ARP(p, r) = \underset{\forall ~(G_{p:v}, G_{p:j}) \in X}{\mathrm{argmax}} \vert FPR(G_{p:v}, r) - FPR(G_{p:j}, r) \vert
    \label{eq.arp}
\end{equation}
}

Finally, to measure how similar rankings are (both base and consensus rankings) we employ the rank similarity Kendall-Tau distance \cite{kendall1938new}. Equation \ref{eq.kt} expressed the Kendall-Tau distance between two rankings $r_1$ and $r_2$.
\begin{equation}
\begin{gathered}
 dist_{KT}(r_1,r_2) = \\
 \vert \{\{x_i, x_j\}\in X: x_i \prec_{r_1} x_j \text{ and } x_j \prec_{r_2} x_i  \}\vert
\end{gathered}
\label{eq.kt}
\end{equation}

\subsection{Consensus Generation}

A {\bf consensus ranking} is a ranking that best represents the preferences of the rankers (i.e., the base rankings). Typically, a consensus ranking is derived by minimizing a distance metric between itself and all the base rankings. Finding a consensus ranking corresponds to the traditional rank aggregation task \cite{brandt2012computational}. Cachel et. al. \cite{cachel2022manirank} augments finding a consensus ranking by addressing group fairness. That is, they provide algorithms for finding a consensus ranking such that $ARP = \delta$, where $delta = [0,1]$. In our system we employ the Fair-Copeland method, please see \cite{cachel2022manirank} for applicable pseudocode. When the system is tasked with finding a consensus ranking (not fairness constrained whatsoever) this is expressed through the fairness threshold where $t = 0$, which represents  $1 - \delta$ or ARP = 1. The consensus ranking returned will maximize representing the preferences of the rankers. However, if the base rankings are unfair, as quantified by the $ARP$ measure, the consensus ranking will also exhibit unfairness. By increasing the fairness threshold the user tunes the resulting fairness (as expressed by lower $ARP$ measures) of the returned fair consensus ranking. 

% Todo caption
\begin{figure*}[h!]
    \centering
    \includegraphics[width=\linewidth]{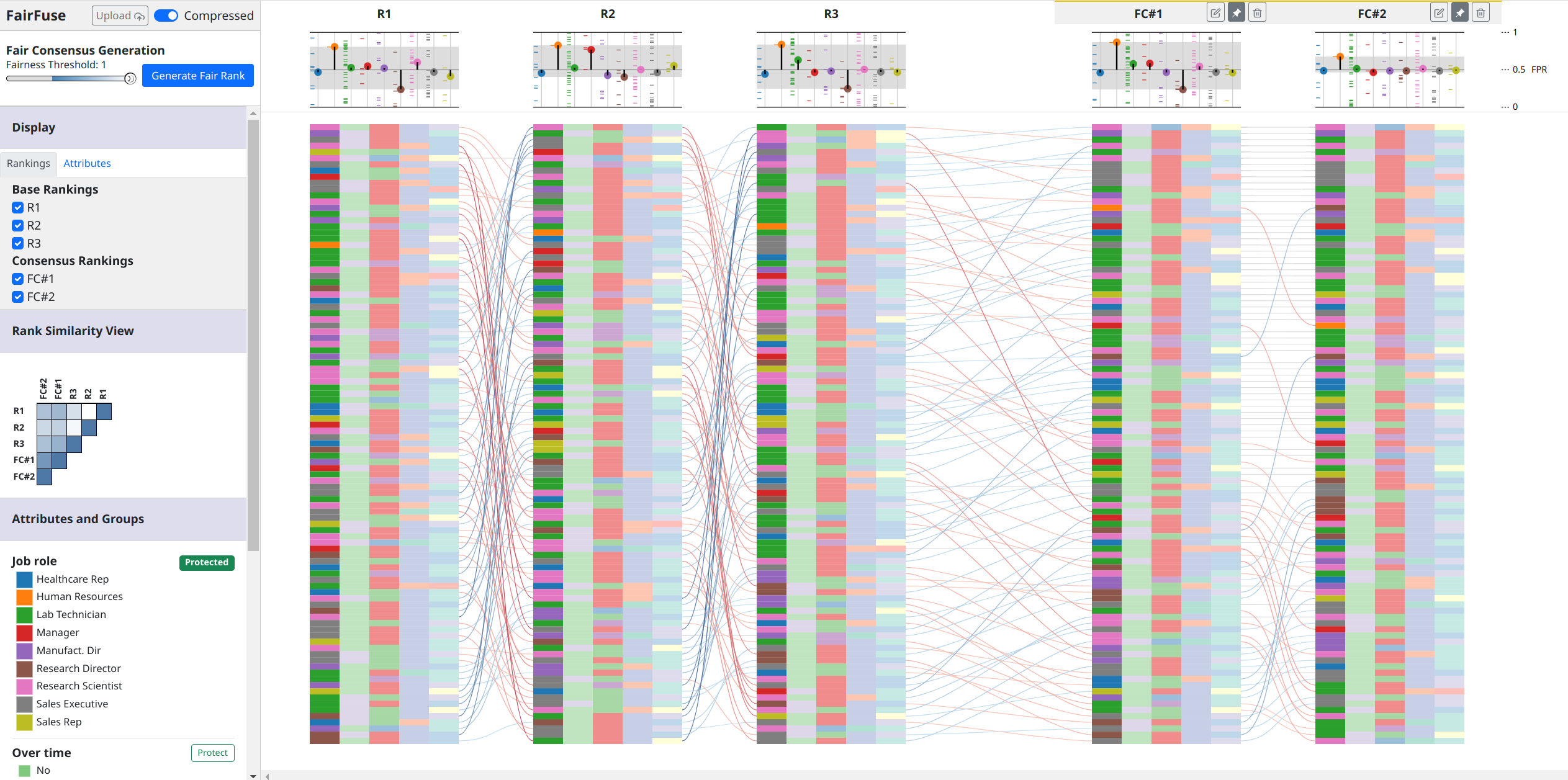}
    \caption{\systemName used for Employee bonus distribution use case. Manager observes disagreement between committee members from the Ranking Exploration View. Bias between different job roles is observed from the Group Fairness View.}
    \label{fig:employee-bonus}
\end{figure*}

\begin{figure*}[h!]
    \centering
    \includegraphics[width=\linewidth]{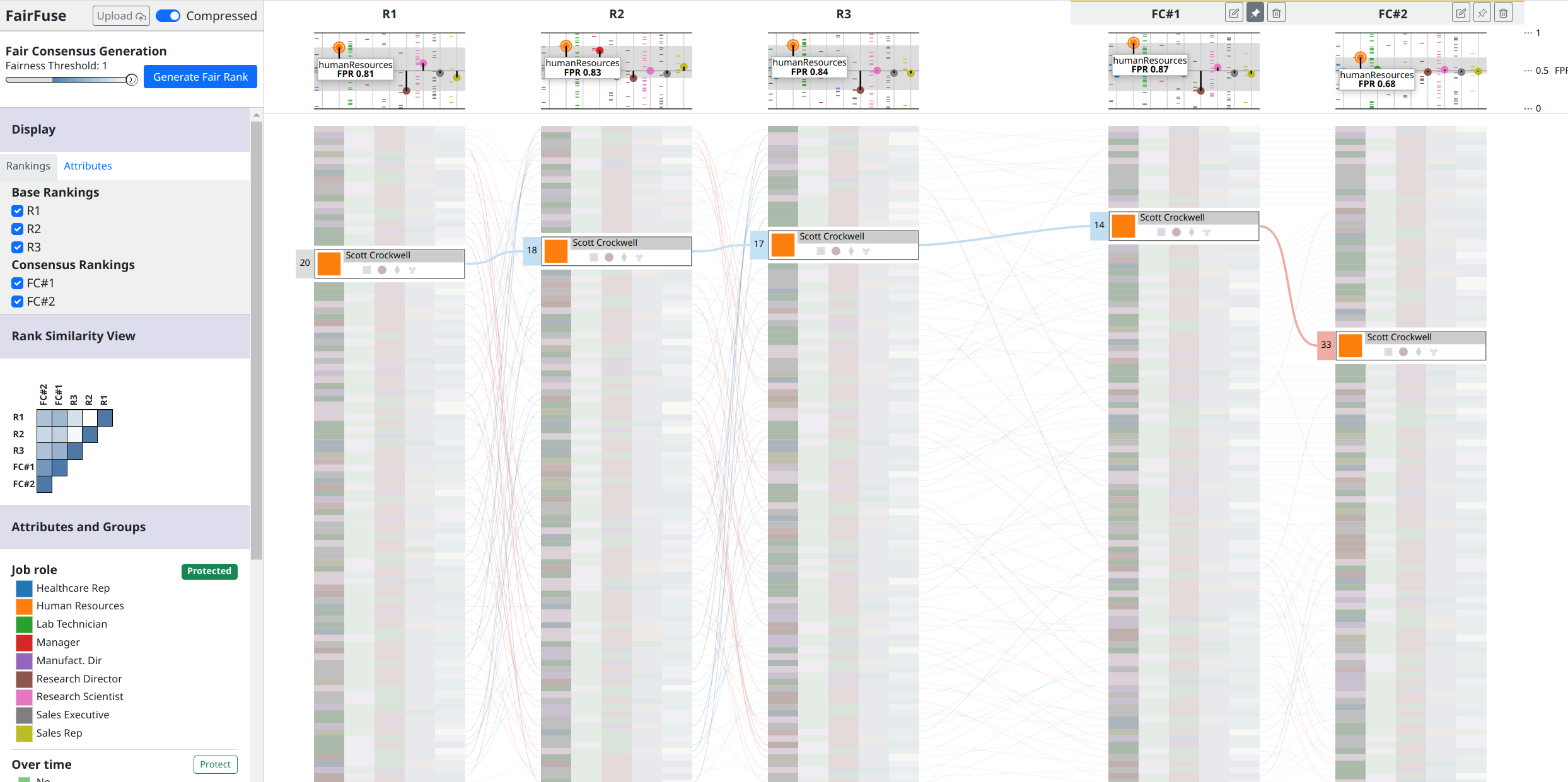}
    \caption{Selecting a group in the Group Fairness View highlights candidates of that group. In this case, there is only single candidate representing Human Resource Group.}
    \label{fig:employee-bonus-group-selection}
\end{figure*}

\section{Sup-3: Use Case Scenario for Employee Bonus Distribution}

A manager is tasked with deciding annual bonuses for employees in their organization. They setup a committee with leads from different departments within the organization who rank all the employees. The manager now needs to combine each of the committee preferences into a final consensus ranking so that bonus awards can be allocated to employees. However, they also want it to be fair with respect to employees from different job roles to not advantage certain role at bonus time. To tackle this challenge, all the rankings made by the committee members are uploaded to the \systemName tool with job roles as a protected attribute.

The manager immediately notices that rankings by the committee have a lot of disagreement by inspecting the number of crossings in the Ranking Exploration View. Looking at the Group Fairness View, they notice that the Human Resource (HR) group is highly advantaged while Research Directors are highly disadvantaged across all three rankings. Upon generating a consensus ranking, the manager notices this bias is transferred and reflected in consensus ranking as well. So, they decide to mitigate bias by adjusting the Fairness Threshold Slider and re-generating the consensus ranking. While the ARP score is significantly reduced, they notice that HR group is still advantaged (Figure \ref{fig:employee-bonus}) with higher FPR score. 

Looking back at the heatmap in the Group Fairness View, there seems to be a single employee as HR. So, to inspect this employee they select the dot that represents HR. The HR employee is highlighted in the ranking exploration view as well. %This helps the employer to investigate more on the particular employee.
They click on the highlighted employee in the ranking exploration view and realize that the employee's rank does not change much between base rankings by the committee members, but the generated fair consensus ranking lowers this employees position (Figure \ref{fig:employee-bonus-group-selection}). The employer does not like this as the employee is the only one as an HR. Hence, they decide to manually adjust the ranking of the employee by dragging the employee card higher in the rank. Happy with the final fair consensus ranking, the manager proceeds with distributing bonuses based on the ranking.

%% if specified like this the section will be committed in review mode
%\acknowledgments{
%The authors wish to thank A, B, and C. This work was supported in part by
%a grant from XYZ (\# 12345-67890).}

\newpage
\clearpage

\onecolumn{
\bibliographystyle{abbrv-doi}

}
\bibliography{supplemental_material_references}